\documentclass[manuscript=article]{achemso}
\usepackage[dvipsnames]{xcolor}
\usepackage{subcaption}
\usepackage{amssymb}
\usepackage{amsmath}
\usepackage{array}
\newcolumntype{L}[1]{>{\raggedright\let\newline\\\arraybackslash\hspace{0pt}}m{#1}}
\newcolumntype{C}[1]{>{\centering\let\newline\\\arraybackslash\hspace{0pt}}m{#1}}
\newcolumntype{R}[1]{>{\raggedleft\let\newline\\\arraybackslash\hspace{0pt}}m{#1}}
\usepackage[version=3]{mhchem}
\usepackage{xr}

\title{Random Forest Models for Accurate Identification of Coordination Environments from X-ray Absorption Near-Edge Structure}
\author{Chen Zheng}
\affiliation[UCSD]{Materials Virtual Lab, Department of NanoEngineering, University of California San Diego, 9500 Gilman Dr, Mail Code 0448, La Jolla, CA 92093-0448, United States}
\altaffiliation{These authors contribute equally to this work.}
\author{Chi Chen}
\affiliation[UCSD]{Materials Virtual Lab, Department of NanoEngineering, University of California San Diego, 9500 Gilman Dr, Mail Code 0448, La Jolla, CA 92093-0448, United States}
\altaffiliation{These authors contribute equally to this work.}
\author{Yiming Chen}
\affiliation[UCSD]{Materials Virtual Lab, Department of NanoEngineering, University of California San Diego, 9500 Gilman Dr, Mail Code 0448, La Jolla, CA 92093-0448, United States}
\author{Shyue Ping Ong}
\affiliation[UCSD]{Materials Virtual Lab, Department of NanoEngineering, University of California San Diego, 9500 Gilman Dr, Mail Code 0448, La Jolla, CA 92093-0448, United States}
\email{ongsp@eng.ucsd.edu}

\date{}

\begin{document}

\maketitle
\newpage
\begin{abstract}
Analyzing coordination environments using X-ray absorption spectroscopy has broad applications ranging from solid-state physics to material chemistry. Here, we show that random forest models can identify the main coordination environment from K-edge X-ray absorption near edge structure (XANES) with a high accuracy of 85.4\% and all associated coordination environments with a high Jaccard score of 81.8\% for 33 cation elements in oxides, significantly outperforming other machine learning (ML) models. In a departure from prior works, we used a robust description of the coordination environment as a distribution over 25 distinct coordination motifs with coordination numbers ranging from 1-12. The random forest models were trained on the world's largest database of $\sim190,000$ computed K-edge XANES spectra. Furthermore, the random forest models can be applied to predict the coordination environment from experimental K-edge XANES with minimal loss in accuracy (82.1\%) due to the use of data augmentation. A drop-out feature importance analysis highlights the key roles that the pre-edge and main-peak regions play in coordination environment identification, with the post-peak region becoming increasingly important at higher coordination numbers. This work provides a general strategy to identify the coordination environment from K-edge XANES across broad chemistries, paving the way for future advancements in the application of ML to spectroscopy.
    
\end{abstract}

\section{Introduction}
X-ray absorption spectroscopy is an important technique for probing the local environments, i.e., atomic coordination symmetries, the number and chemical identities of neighboring atoms and oxidation states, in a material.\cite{odayXRayAbsorptionSpectroscopy2000, chaurandNewMethodologicalApproach2007, silversmitStructureSupportedUnsupported2005} The X-ray absorption spectra (XAS) consists of the X-ray absorption near-edge structure (XANES) at low energy and the extended X-ray absorption fine structure (EXAFS) at high energy. While quantitative analysis of the EXAFS is relatively mature, analysis of the XANES is challenging due to its sensitivity to many factors including coordination number (CN)\cite{fargesTiEdgeXANES1997, fargesTransitionElementsWaterbearing2001}, orbital hybridization\cite{debeergeorgeMetalLigandKEdge2005}, spin state\cite{westreMultipletAnalysisFe1997}, oxidation state\cite{yamamotoAssignmentPreedgePeaks2008} and symmetry\cite{sanoXANESSpectraCopper1992} of the central absorbing atoms. However, the XANES signal usually dominates the XAS spectrum and, in principle, provides richer information regarding the coordination environments compared to EXAFS. 

A typical analysis of XANES relies on comparisons between experimentally measured spectroscopy and spectra from well-known compounds.\cite{fargesPreedgeAnalysisMn2009, fernandez-garciaXANESAnalysisCatalytic2002}. There have been attempts for quantitative interpretations of XANES spectra using principal component analysis\cite{manceauEstimatingNumberPure2014, fayDeterminationMoSurface1992,beaucheminPrincipalComponentAnalysis2002} and linear deconvolution methods\cite{bajtXrayMicroprobeAnalysis1994}. These approaches seek to break down the XANES spectrum of a multi-component system into individual component spectra, which provide the statistical basis for estimating the presence and ratios of individual species. However, these techniques are difficult to apply to systems that do not have well-established reference spectra. Theoretical calculations based on time-dependent density-functional theory (TDDFT)\cite{tanakaFirstprinciplesCalculationsXray2009}, multi-scattering\cite{rehrTheoreticalApproachesXray2000,rehrParameterfreeCalculationsXray2010}, and Bethe-Salpeter equation (BSE) approaches\cite{laskowskiUnderstandingXrayAbsorption2010} provide an alternative means of obtaining the XANES of any material. Recently, the current authors have developed the first-of-its-kind large, public database of X-ray absorption spectra (XASDB).\cite{zhengAutomatedGenerationEnsemblelearned2018,mathewHighthroughputComputationalXray2018} Based on the FEFF multi-scattering code\cite{rehrParameterfreeCalculationsXray2010}, 580,000 K-edge XANES spectra of over 52,000 crystals in the Materials Project have been calculated and are freely available in the XASDB at the time of writing.\cite{jainCommentaryMaterialsProject2013} This database not only provides an important reference for experiments but also opens new paths for large-scale quantitative XANES analysis. For example, the authors have previously shown that an ensemble-learning spectra matching algorithm can achieve a 84.2\% accuracy in identifying oxidation state and local environment by matching unknown spectra with computed spectra in the XASDB.\cite{zhengAutomatedGenerationEnsemblelearned2018}

The extraction of coordination environment information from the XANES is akin to that of image recognition, a field where ML techniques have made great strides. Indeed, there have been attempts to apply ML to quantitative and qualitative XANES analysis. For example, \citet{timoshenkoSupervisedMachineLearningBasedDetermination2017} have demonstrated that neural networks can predict the CN of Pt atoms from L-edge XANES spectra of metallic nanoparticles. \citet{carboneClassificationLocalChemical2019} have also shown that convolutional neural networks (CNNs) can predict the coordination environments of 3$d$ transition metal species from site-specific K-edge XANES spectra. The accuracy achieved was an impressive 86\%. However, the work has focused on three types of well-defined coordination, i.e., tetrahedral, square pyramidal, and octahedral, and as acknowledged by the authors themselves, the dominant octahedral environment makes up 64\% of the total data. In addition, previous works have reported that material information, such as chemical, elemental and geometric information, can be obtained from the interpretation of calculated oxygen K-edges ELNES/XANES spectra of metal oxides and \ce{SiO2} using decision tree methods.\cite{kiyoharaDatadrivenApproachPrediction2018} Very recently, \citet{suzukiAutomatedEstimationMaterials2019} have used L-edge XANES or EELS spectra of MnO in conjunction with a regression model to capture the crystal-field parameters.

Despite these advances, two crucial gaps remain. The main limitation is that previous works have treated coordination environment identification as a classification problem between mutually-exclusive labels. In reality, the coordination environment can be represented along a continuum. For instance, when a species in a perfect regular octahedron is displaced towards one of the vertices, its coordination environment becomes increasingly square pyramidal-like but still retains features of octahedral coordination. A rigorous treatment of coordination environment, therefore, needs to define how ``square pyramidal-like'' and ``octahedron-like'' the coordination environment is. A second major limitation is that previous works focus either on a very narrow set of chemistries or environments using experimental XANES data\cite{timoshenkoSupervisedMachineLearningBasedDetermination2017} or a somewhat broader set of chemistries and environments using computed XANES data only\cite{carboneClassificationLocalChemical2019}. Given the well-known errors in computed lattice parameters and XANES, it is unclear how ML models trained on large and diverse computed XANES can be applied to experimental XANES.

\begin{figure}[ht!]
    \centering
    \includegraphics[width=0.9\textwidth]{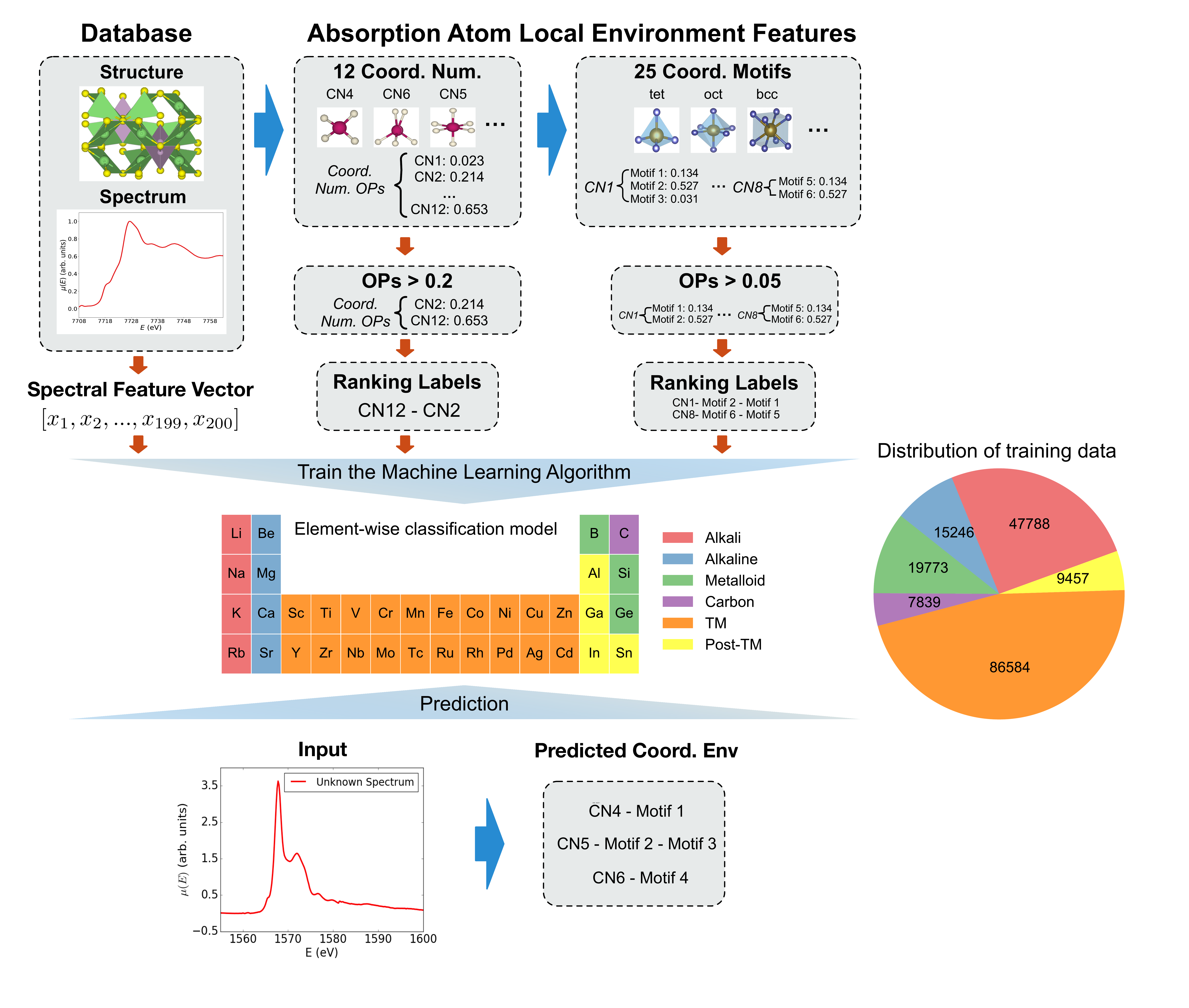}
    \caption{Workflow schema of the coordination environment identification algorithm.}
    \label{fig:workflow}
\end{figure}

In this work, we comprehensively address the above limitations and develop an approach to identify local environments in oxides from K-edge XANES using random forest models. In contrast to prior models, CNs up to 12, and a total of 25 distinct coordination motifs (CMs), which are enumerated in Figure S1 are considered. The model accuracy is assessed by correctly predicting the ranking of the coordination environments with their probabilities above a certain threshold, for example, predicting a six-coordinated atom to have $octahedral$, $pentagonal\,pyramidal$ and $hexagonal\,planar$, in decreasing probability. This is a much more comprehensive yet difficult problem to solve than predicting a single CM since as correctly predicting the dominant CM (e.g., $octahedral$) may still render a false prediction if the correct order for secondary CMs is wrong. High prediction accuracy of $\sim 85.4\%$ was achieved over 33 cations in oxides, covering most technologically-relevant cation species including alkali, alkaline, metalloid, transitional metals, post-transition metals and carbon (see Figure \ref{fig:workflow}). Most importantly, we demonstrate the augmentation of the training data with broadened/compressed spectra to mimic the effect of DFT lattice parameter prediction error on spectra. The resulting models can be directly applied to identify coordination environments from experimental XANES with minimal loss of accuracy.

\section{Results}

\subsection{Dataset construction}

The training data was constructed from a database of $\sim190,000$ site-specific K-edge XANES of $\sim22,500$ oxides in the Materials Project.\cite{zhengAutomatedGenerationEnsemblelearned2018,mathewHighthroughputComputationalXray2018,jainCommentaryMaterialsProject2013} To the authors' best knowledge, our dataset represents the broadest coverage of cation elements to date in the study of XANES. Figure \ref{fig:workflow} provides a summary of the total dataset used in this work. Cation elements with atomic number larger than 52 were excluded due to the lack of distinguishable K-edge spectral features. From each spectrum, an energy window of 45 eV from the spectral onset was extracted and converted to a vector of 200 intensity values using linear interpolation. This is the strong scattering XANES region covering the pre-edge, main-peak and post-peak spectral features.\cite{ankudinovRealspaceMultiplescatteringCalculation1998} All three regions have shown to be critical for the identification of local coordination environments.\cite{carboneClassificationLocalChemical2019} The intensity vector was then normalized so that the value of maximum magnitude equals 1. The experimentally-measurable element-wise spectra, i.e., the average of all spectra for a particular absorbing element in a structure, were also included in the training data.

In our previous work\cite{zhengAutomatedGenerationEnsemblelearned2018}, we have found that the broadness of the computed XANES  feature is sensitive to the lattice parameter variation. To improve the robustness of the classification models, the initial dataset was augmented by randomly sampling 30\% of spectra and applying broadening or compression of $\pm5$eV in energy range to mimic the variations in feature broadness. This spectral shape distortion corresponds to up to $7\%$ variations in the lattice parameters, which exceeds the $\sim$5\% systematic errors introduced by the Perdew-Berke-Ernzerhof (PBE)\cite{perdewGeneralizedGradientApproximation1996} generalized gradient approximation function used in the Materials Projects for crystal structure optimization. 

For each site, the coordination environment is defined as the combination of the CN and the CM. Figure S1 provides a comprehensive enumeration of the CMs considered in this work. Coordination environment determination for a known structure was carried out using the algorithm by \citet{zimmermannAssessingLocalStructure2017}, as implemented in pymatgen\cite{ongPythonMaterialsGenomics2013} and matminer\cite{wardMatminerOpenSource2018}. The algorithm consists of two steps. The first step identifies the number of bonded neighbors to an atom based on the Voronoi tessellation method. The solid angle weights of all neighbors are used to determine a site CN order parameter (OP) that describes how consistent a site is with a certain CN. The CN OPs values range from 0 to 1, with 1 representing perfect resemblance. An OP vector $\vec{p}$ is constructed for each site for CNs ranging from 1 to 12, as follows.
\begin{equation}
    \vec{p} = \{p_1, p_2, p_3, p_4, ..., p_{12}\}, \mathrm{where} \sum_{i=1}^{12} p_i = 1
\end{equation}
where $p_i$ denotes the OP for a CN of $i$. CNs greater than 12 are not considered due to their extremely low counts in the data set, as shown in Figure S2. $\vec{p}$ is a more robust statistical representation of a CN compared to using a single CN value. For example, a site may have $p_4 = 0.2$ and $p_6 = 0.8$, indicating that it mostly resembles a CN of 6 and shares some similarity to a CN of 4. This is in contrast to a single-valued CN that is sensitive to radius cutoffs used to determine neighbors and classification. In practice, the CN labels are generated by setting a cutoff for $p_i$ and then concatenating the probability-sorted CNs (see Figure \ref{fig:workflow}). In the second step, the CM is determined by matching the neighbors identified in the first step to prototype motifs. For example, six-fold coordination can result from hexagonal planar, octahedral and pentagonal pyramidal coordination. Again, a vector of OPs $\vec{q}$ based on twenty-five prototype motifs is computed for each site, as follows:

\begin{eqnarray*}
        \vec{q}  = & \{ & q_{single\, bond}\times{p_1}, ..., q_{tetrahedron}\times{p_4}, q_{octahedron}\times{p_6},\\ & & q_{hexagonal\,planar}\times{p_6}, q_{pentagonal\,pyramidal}\times{p_6}, ..., q_{cuboctahedra}\times{p_{12}}\},
\end{eqnarray*}

where $q_i$ denotes the OP for a CM prototype of $i$. The CN OPs are factored into the vector of CM OPs $\vec{q}$. The CMs are not mutually exclusive, and hence, their OP sum will not be 1.  In this step, we did not consider CN9, CN10 and  CN11 since they do not have dedicated CMs. Similarly, the CM labels are generated by setting a threshold for CM and concatenating the probability-sorted CMs, as shown in Figure \ref{fig:workflow}. Our strategy of using ranking labels provides a rich representation of the coordination environment. The ranking labels of CM OPs were encoded for a specific type of CN. For example, we took into account only $\{q_{octahedron}\times{p_6}, q_{hexagonal\,planar}\times{p_6}, q_{pentagonal\,pyramidal}\times{p_6}\}$ for generating CM ranking label of CN $= 6$ (see Methods section for details). 

The coordination environment classification task can then be divided into two sequential steps powered by two separate models for each element. In the first step, the CN model identifies the CN ranking label from the spectra, and in the second step, the CM model identifies the CN-specific CM ranking label. The models are trained for each element as the characteristic XAS absorption edge energy follows a power law with atomic number and is well separated\cite{newvilleFundamentalsXAFS2014}. The absorbing species can be identified with 100\% accuracy from simply examining the spectral energy range. This domain knowledge significantly reduces the problem complexity and is expected to improve model accuracy. Eventually, the coordination environment recognition problem becomes a two-step multi-label classification problem, where an absorption spectrum might reflect a statistical ensemble of more than one coordination environment. This is an attractive problem transformation approach which provides both scalability and flexibility\cite{readClassifierChainsMultilabel2011} to handle most off-the-shelf multi-label classification algorithms\cite{changLIBSVMLibrarySupport2011,breimanRandomForests2001,pedregosaScikitlearnMachineLearning2011}. 

\subsection{Machine learning models}

Figure \ref{fig:workflow} provides an overview of the coordination environment classification workflow. As some elements are found only in specific local environments\cite{waroquiersStatisticalAnalysisCoordination2017}, the knowledge of elemental types would already significantly narrow the range of possible local environments. Indeed, a ``baseline'' model can be constructed that merely assigns a CN-CM classification based on the dominant environment for that element. Such a baseline model has a high classification accuracy of 70-80\% on the first row transition metal cations from Sc to Ni, an intermediate accuracy of $\sim$ 60\% for the post-transition metals and metalloid, and a relatively low accuracy of 17-58\% for the alkali and alkaline earth cations (see Figure \ref{fig:acc_improvement}). Any reasonable ML model, therefore, has to achieve a substantial improvement over this ``baseline'' model across all chemical classes.

In the next steps, optimized element-specific ML models sequentially identify firstly the CN ranking label, followed by the CN-specific CM ranking label, from the spectra. Five ML models were assessed in terms of the performance in CN and CM classification, namely $k$-nearest neighbor ($k$NN), random forest, multi-layer perceptron (MLP)\cite{lecunDeepLearning2015}, CNN\cite{lecunConvolutionalNetworksApplications2010} and support vector classifier (SVC).  Model fitting and hyperparameter optimization used a five-fold cross validation method. During the optimization process, we performed a grid search to identify optimal values for key ML parameters that are directly related to the classifiers' performances. These parameters include $k$ in the $k$NN model, number of trees in the random forest model, number of neurons/layers and choice of activation function in MLP and CNN, and the penalty parameter $C$ and the kernel coefficient ($\gamma$) for the SVC. For all the other parameters, we used the defaults within the scikit-learn package\cite{pedregosaScikitlearnMachineLearning2011}. Previous works have shown that the performance of the CNN-based model in the classification of XAS spectra is invariant across different neural network structures.\cite{carboneClassificationLocalChemical2019} The same hyper-parameter space was adopted in the optimization of ML models for each classification sub-task (see Methods section for the details). 

As shown in Figure \ref{fig:workflow}, this work focuses only on elements in rows 2-5 of the periodic table, excluding the noble gases; elements in row 6 and beyond, including the rare earth elements, were not investigated because the lack of resolution in the K-edge absorption spectra for elements with atomic number greater than 52. 

\subsection{Computational spectra classification performance}

\begin{figure}[!ht]
\centering
\begin{subfigure}[b]{0.45\textwidth}
\centering
\includegraphics[width=\textwidth]{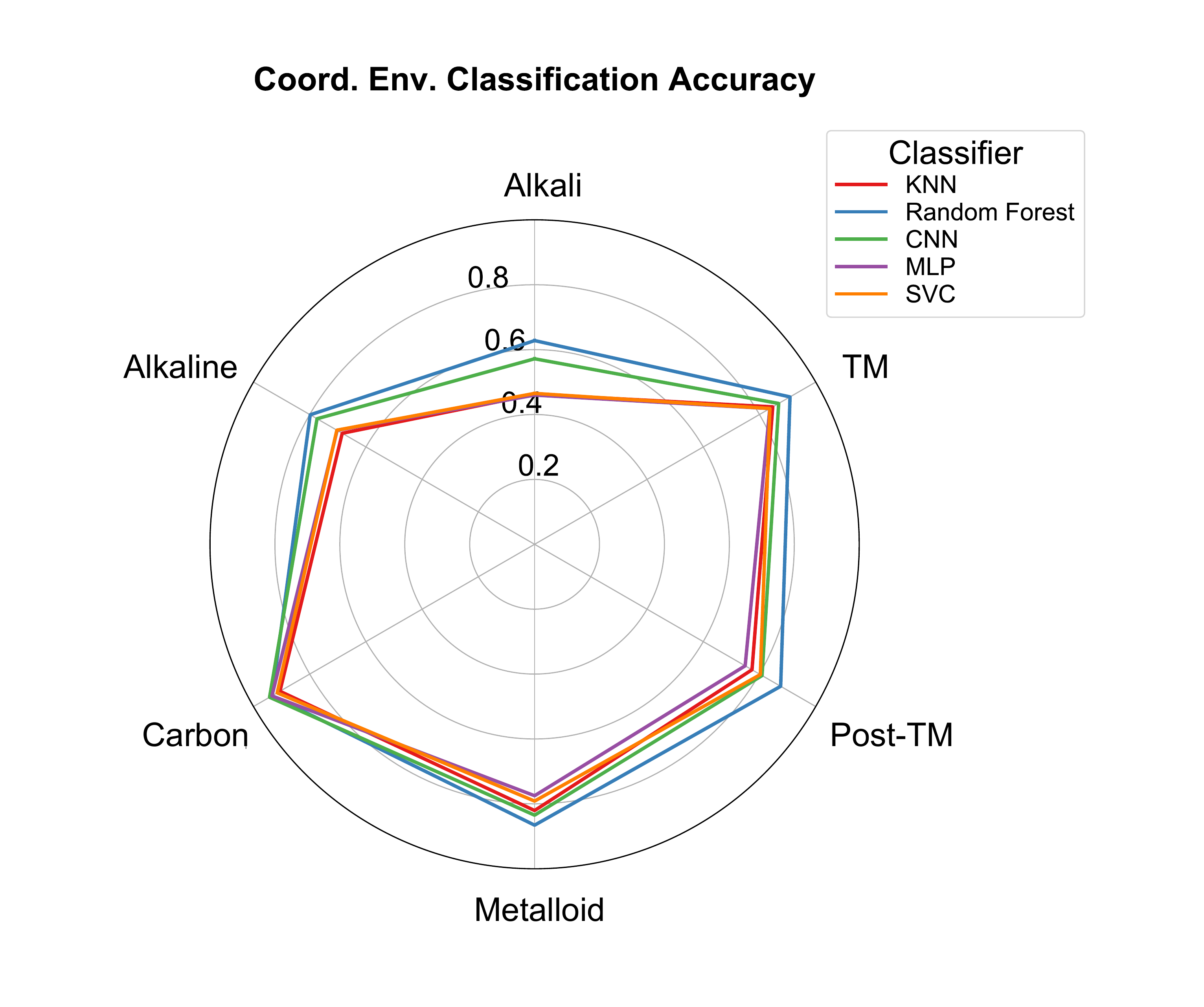}
\caption{Accuracy}
\label{fig:c_num_acc}
\end{subfigure}
\hfill
\begin{subfigure}[b]{0.45\textwidth}
\centering
\includegraphics[width=\textwidth]{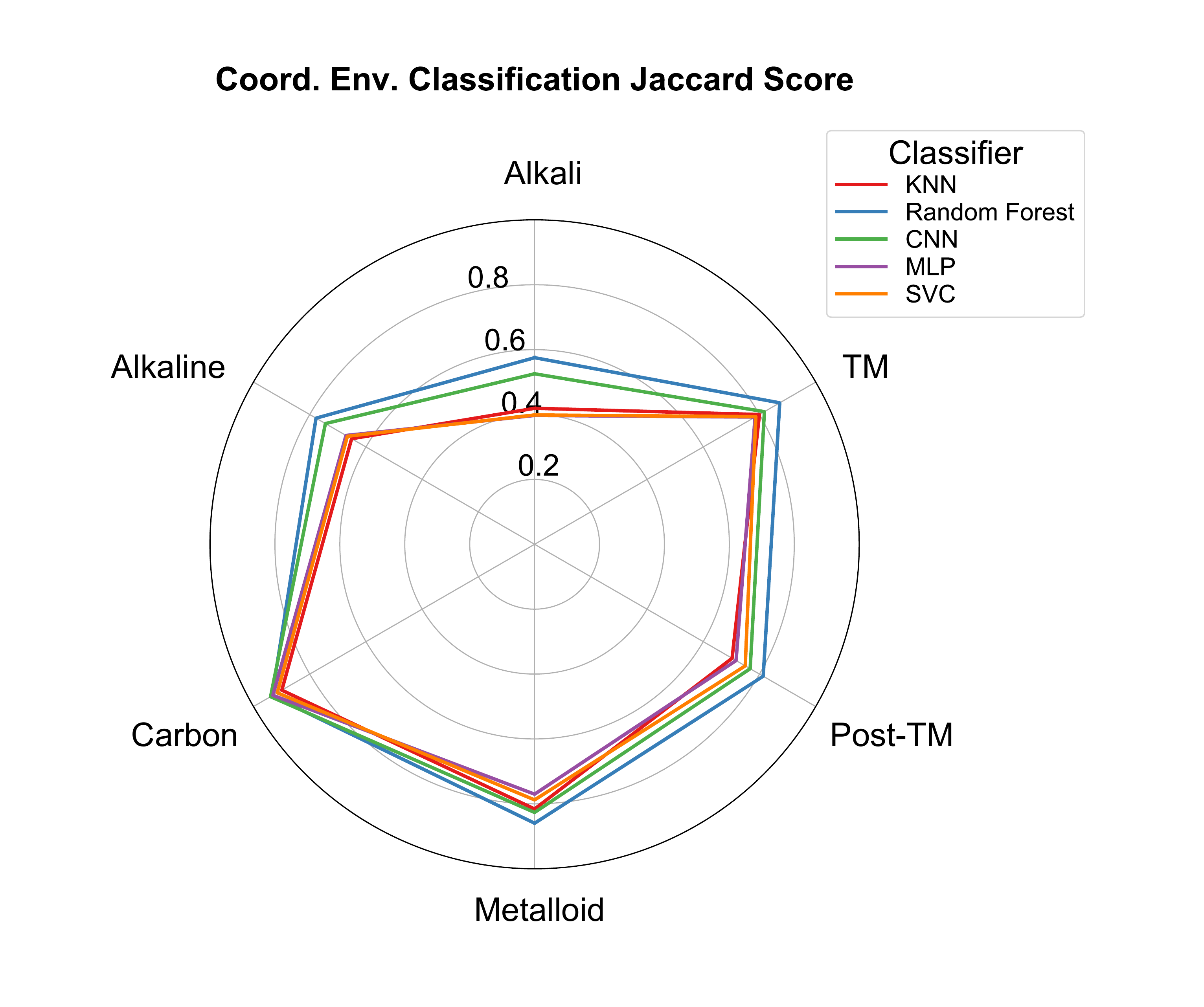}
\caption{Jaccard score}
\label{fig:c_env_jaccard}
\end{subfigure}
\begin{subfigure}[b]{0.4\textwidth}
\centering
\includegraphics[width=\textwidth]{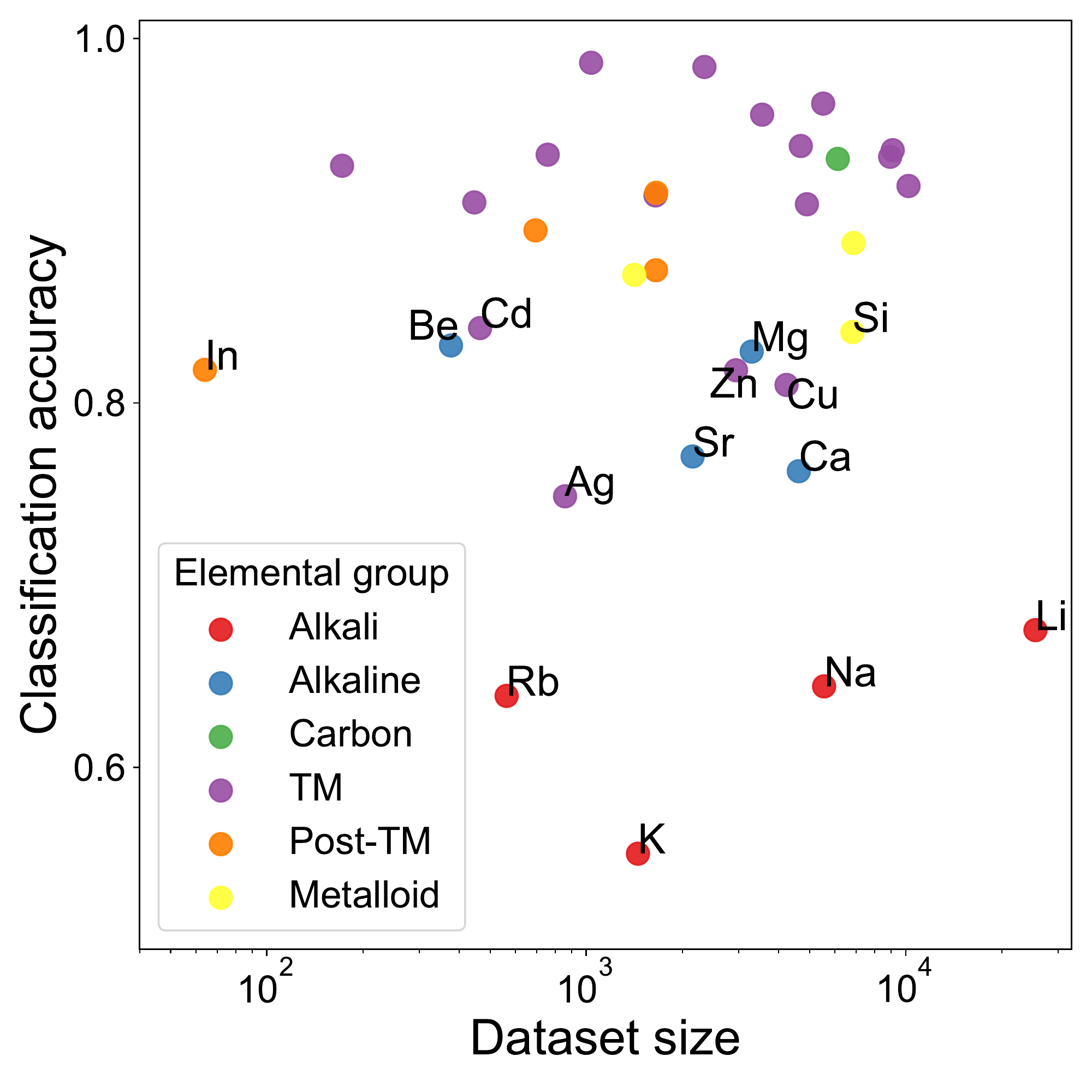}
\caption{}
\label{fig:coord_env_acc_dataset_size}
\end{subfigure}
\hfill 
\begin{subfigure}[b]{0.4\textwidth}
\centering
\includegraphics[width=\textwidth]{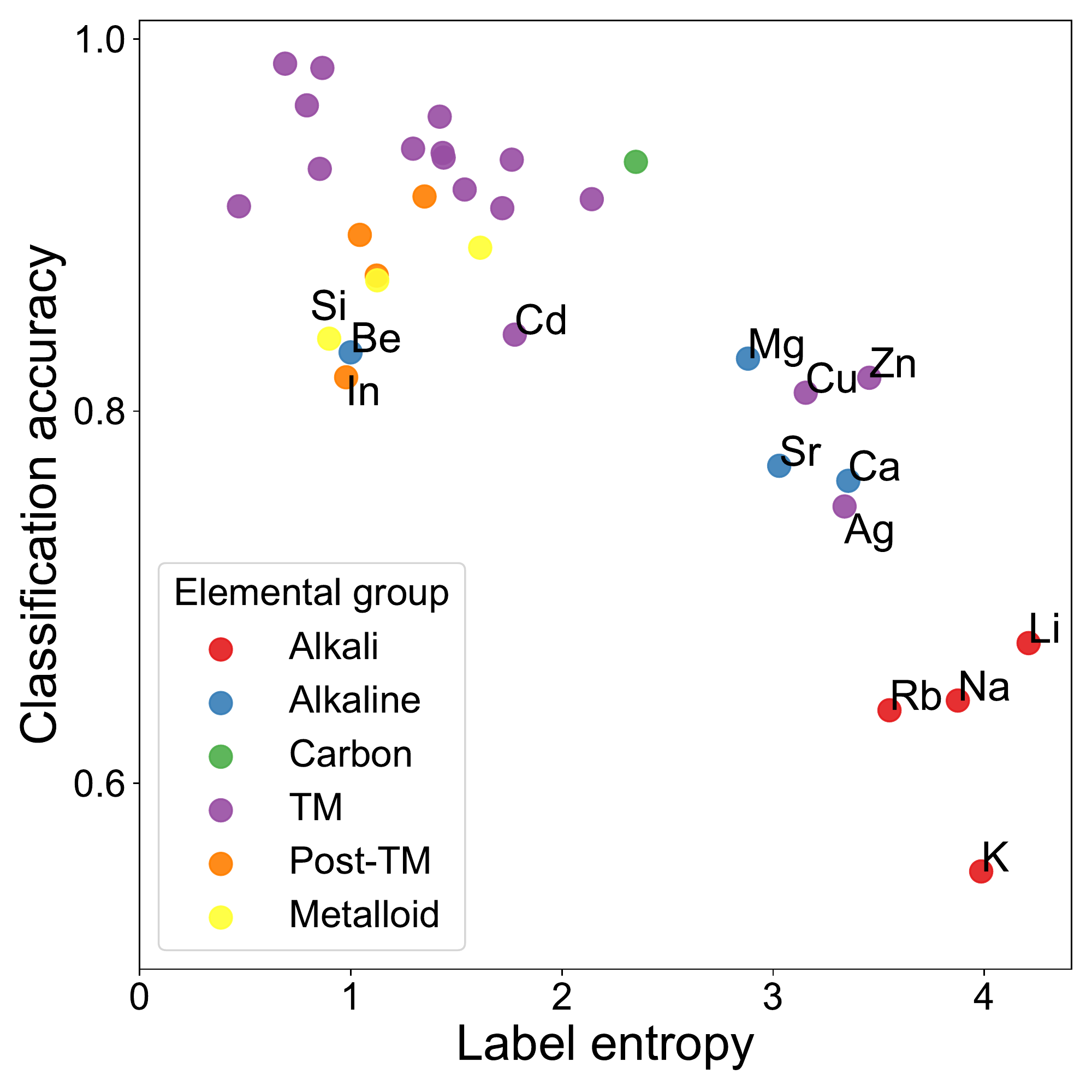}
\caption{}
\label{fig:coord_env_acc_entropy}
\end{subfigure}
\caption{Performance of five ML classifiers - $k$ nearest neighbor ($k$NN), random forest, CNN, multi-layer perceptron (MLP) and support vector classifier (SVC) - on coordination environment classification. (a) Accuracy and (b) Jaccard score for the five ML classifiers broken down by elemental categories, namely alkali metals, alkali earth metals, transiton metals (TM), post-transition metals, metalloids and carbon (see \ref{fig:workflow} for color-coded categories). (c) Relationship between the random forest model's classification accuracy and the dataset size. (d) Relationship between the random forest model's classification accuracy and the training label entropy. Cation elements with classification accuracy less than 0.85 were tagged in Figures \ref{fig:coord_env_acc_dataset_size} and \ref{fig:coord_env_acc_entropy}.}
\label{fig:radar_figures}
\end{figure}

Figures \ref{fig:c_num_acc} and \ref{fig:c_env_jaccard} compare the accuracy and Jaccard index (see Methods for definitions), respectively, of the optimized five classifiers broken down into the six elemental categories. The accuracy captures how well each ML model performs in predicting the top-ranked coordination environment, i.e., the combined CN-CM score with the highest value. The Jaccard index, on the other hand, captures how well each ML model performs in identifying all relevant coordination environments related to the absorbing species, i.e., all CN and CM with non-zero OPs. For all element categories, the random forest classifiers outperform the other classifiers, with an overall accuracy of 85.4\% and a Jaccard score of 81.8\%. 

One key observation from Figures \ref{fig:c_num_acc} and \ref{fig:c_env_jaccard} is that classification performance is highly dependent on elemental category. While the performances of all classifiers are relatively high ($>90\%$ accuracy) for carbon, the performances on the alkali metals are comparatively poor. To elucidate the origin of the performance variations, we have plotted the classification accuracy for the best performing random forest model against training data set size and label entropy in Figures \ref{fig:coord_env_acc_dataset_size} and \ref{fig:coord_env_acc_entropy}, respectively. Here, the label entropy,\cite{shannonMathematicalTheoryCommunication1948} which is an informational measure of the diversity of the coordination environment labels in each elemental category, is computed using the following expression:
\begin{equation}
    S = - \sum_{i}P_{i}log_{2}P_{i},
\end{equation}
where $P_{i}$ is the probability of a ranking label $i$ out of all ranking labels. The label entropy $S$ is high if the variability of the label values is high, i.e., an element exists in a spectrum of coordination environments with similar probabilities. For example, the alkali metals Li, Na and K have high label entropy because they exist in a variety of local environments - tetrahedral, octahedral - with relatively high probabilities, while the transition metals have low label entropy because they exist mainly in the octahedral coordination, with the exception of the higher oxidation states of V and Cr which nearly almost exists in tetrahedral coordination.\cite{waroquiersStatisticalAnalysisCoordination2017} The Jaccard index with data size and label entropy is shown in Figure S3, which shows a similar trend as accuracy. 

From Figure \ref{fig:coord_env_acc_dataset_size}, it may be observed that there is no clear relationship between classifier performance and training data set size. However, a clear inverse relationship between classifier performance and the label entropy can be seen in Figure \ref{fig:coord_env_acc_entropy}. These observations suggest that data size is not the dominating factor, and the current data size for each element seems sufficient to reach convergent results. The decrease in performance with an increase in label entropy is expected, given that it is much more challenging for a classifier to distinguish between several equi-probable environments as opposed to identifying a single dominant label. The especially poor performance on the light alkali elements (Li, Na and K) may already be attributed to the well-known issues with the FEFF software in reproducing the spectra of light alkali elements.\cite{pradoSodiumEdgeXANES2005} In FEFF, the approximation of core-hole potentials can result in too strong screening effects for the core-hole, which causes a tendency to underestimate the white line intensity of light elements in compounds. This might help to explain why among alkali elements, the increase in training dataset size generally leads to an increase in the classification accuracy since the models require more data to accurately model the correlations between spectral features and coordination environments. For example, the label entropy values of all three light alkali cation elements were all close to 4, while their dataset sizes differ greatly. The training dataset size (25,450) of Li is one magnitude higher than the training dataset size (1,451) of \ce{K}, and the classification accuracy of \ce{Li} is 0.12 higher than \ce{K}. For alkaline earth metal (Be, Mg, Ca, Sr), the coordination environment becomes more diverse as the ionic radius increases, and performance drops accordingly. In the dataset, \ce{Be^{2+}} is always four-coordinated while \ce{Mg^{2+}}, \ce{Ca^{2+}}, and \ce{Sr^{2+}} are found to be four-, five-, six-, seven-, and eight-coordinated. 

As a comparison, Figure S4 shows CNN's prediction accuracy as a function of label entropy values. The CNN classifier fails to deliver classification performances comparable to the random forest classifier. This can be attributed to the relatively small data size per element-CM, with an average of $\sim$ 110, Figure S5, since it is known that neural networks-based models generally need more data to train. Unsurprisingly, CNN model performance shows a more notable positive relationship with the data size (see Figure S4b). In addition, the CNN classifier shows a greater decrease in prediction accuracy as label entropy increases. 

\begin{figure}[ht!]
    \centering
    \includegraphics[width=1\textwidth]{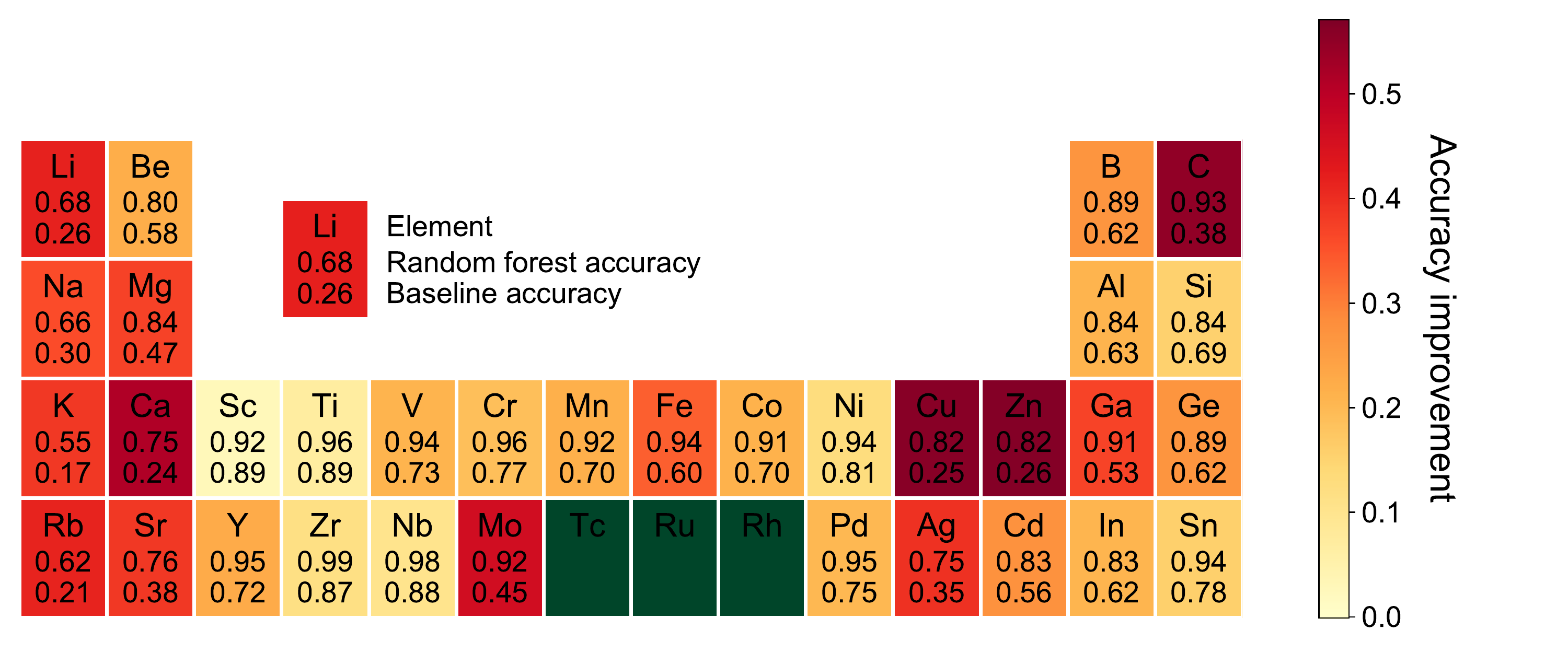}
    \caption{Comparison of accuracy of optimized random forest models with the baseline model for all elements studied. In general, the random forest models outperform the baseline model by significant margins (color of rectangles indicate the level of improvement). Tc, Ru and Rh are excluded due to the lack of data.}
    \label{fig:acc_improvement}
\end{figure}

Figure \ref{fig:acc_improvement} shows a comparison of the accuracy of the random forest models with the ``baseline'' models. The accuracy of the random forest models are well over 80\% for the majority of elements and exceeds 55\% even in the more challenging alkali elements. In general, the random forest models far outperform the ``baseline'' models. High Jaccard indexes are also achieved across the periodic table, as shown in Figure S6.

\subsection{Coordination environment identification from experimental XANES spectra}

We evaluated the random forest classifiers using 28 high-quality normalized XANES experimental spectra, obtained from the XAFS Spectra Library\cite{XASSpectraLibrary},  EELS database\cite{ewelsCompleteOverhaulElectron2016}, supplemented by six high-quality experimental XANES spectra of \ce{V2O5}, \ce{V2O3}, \ce{VO2}, \ce{LiNiO2}, \ce{LiCoO2}, and \ce{NiO} from previous studies.\cite{ranaLocalStructuralChanges2014,ranaStructuralIntegrityElectrochemical2014} These 28 spectra comprise a diverse dataset covering 13 chemical species for classifiers' performance assessment. For spectra from the EELS database and XAFS Spectra Library without available structural information, we assumed that they correspond to the ground state structures in the Materials Project database with the same chemical composition.

We selected the spectral region from -5 eV to 55 eV with reference to edge energy ($E_{0}$) determined by the MBACK algorithm\cite{wengMethodNormalizationXray2005}. As PBE usually leads to up to 5\% lattice parameter overestimation error,\cite{wuMoreAccurateGeneralized2006, haasCalculationLatticeConstant2009} the expanded spectral region encompasses this artificial spectral feature difference between computational and experimental XANES spectra. It should be stressed, however, that the experimental spectra were not used in the training of the random forest models.

The random forest classifier successfully identified 23 of 28 top coordination environment ranking labels, with a coordination environment prediction accuracy of 82.1\% and a Jaccard score of 80.4\%. These accuracies are comparable with those achieved on the computational test set. The random forest classifiers failed to predict the correct coordination environment for two phases of \ce{V2O5}, \ce{ZnO}, \ce{Na2O} and \ce{CuO} spectra, although the models predicted the dominant CN (CN with highest $p_{CN}$) with 100\% accuracy. For \ce{V2O5}, the classifier successfully predicts the dominant CM, i.e., trigonal bipyramidal, but does not predict the correct order of secondary and tertiary CMs (a failure by our strict definition). The likely reason for this failure is the small difference in OPs between the second (i.e., $q_{pentagonal\,planar}$) and third (i.e., $q_{square,pyramidal}$) ranked CMs of $\sim0.029$. In \ce{ZnO}, the coordination environment of Zn does not resemble any target CMs, i.e., all CM OPs are $< 0.22$. Here, the relatively low resemblance between the absorbing atom's coordination pattern and target motifs seems to be the critical issue. For \ce{Na2O}, the failure of the model may be attributed to the possible contamination of the experimental sample.\cite{zhengAutomatedGenerationEnsemblelearned2018} Finally, for \ce{CuO}, the \ce{Cu^{2+}} has a four-fold coordination with oxygen that is matched with five target motifs. The OPs of three of the matched CMs - $rectangular\,see{-}saw{-}like$, $see{-}saw{-}like$ and $square\,co{-}planar$ - exceed 0.5. In this case, the use of EXAFS may be required to identify the local environment with sufficient resolution. 

\subsection{Model insights}

We performed feature importance analysis to gain insights into the contribution of different regions of the K-edge XANES spectra to coordination environment information.  The studied cases include CN = 2 - 8 for all 33 elements in this work. We divided each K-edge XANES spectra into three regions: the pre-edge, main-peak and post-peak with energy range 0-15 eV, 15-30 eV and 30-45 eV, respectively, referenced to the spectral onset. A robust brute-force drop-variable importance approach was used, where part of the input features was systematically dropped to assess the change in model prediction accuracy. In principle, dropping more important features will lead to poorer model performance. The advantage of the drop-variable importance measure is that it provides the ground truth feature importance compared to alternative importance measures\cite{stroblBiasRandomForest2007}. Both single and combined regions, i.e., ``Pre+Main'', ``Pre+Post'' and ``Main+Post'', were also investigated.

\begin{figure}[htp]
\centering
\includegraphics[width=1\textwidth]{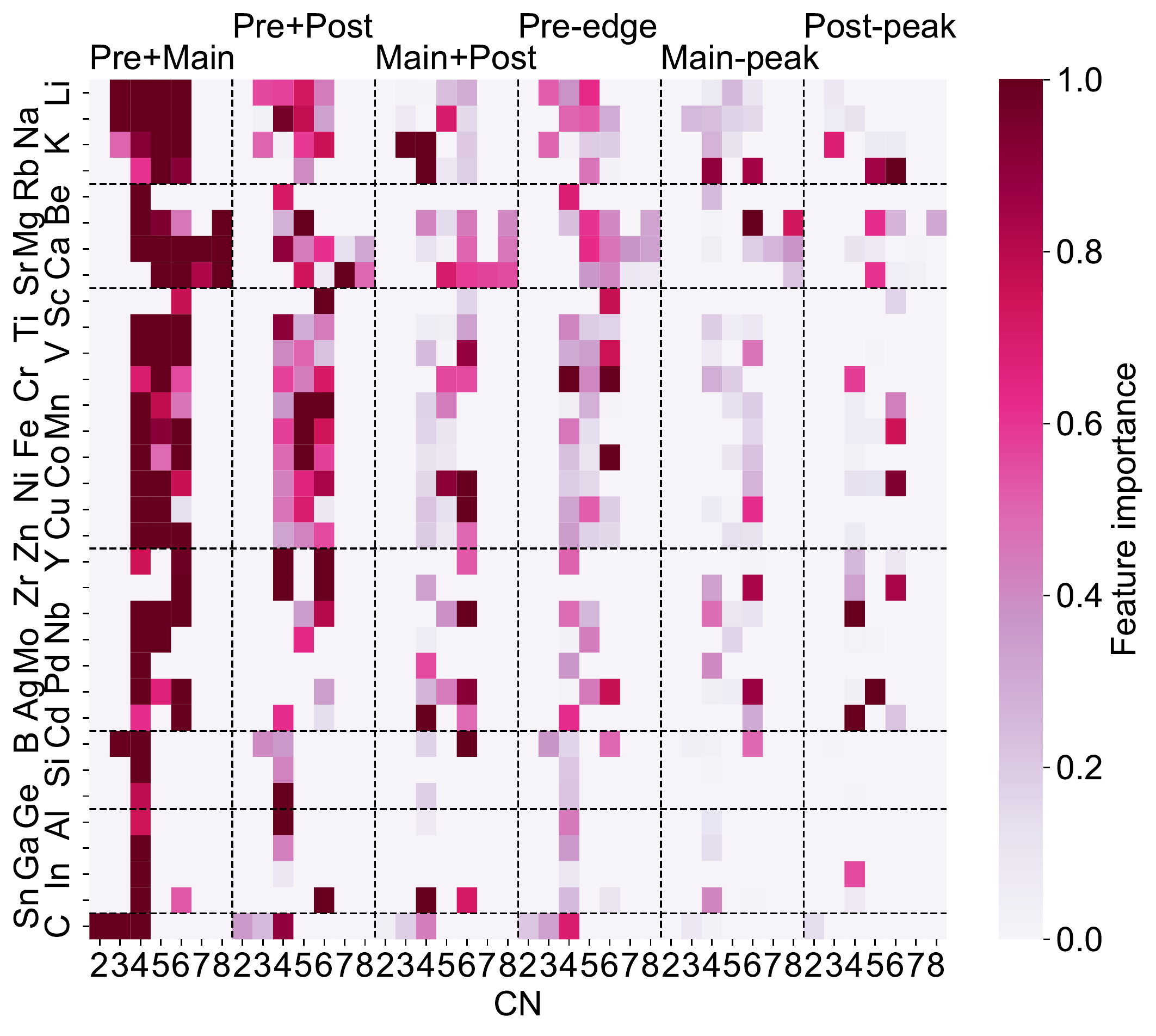}
\caption{Normalized feature importance of different regions of spectra for predicting a given CN for each element. The drop-variable feature importance is normalized with respect to the maximum importance of a spectral region for each element. The x-axis is arranged by spectral regions (i.e., pre+main, pre+post, etc.) followed by increasing CN within each spectral region. The y-axis is arranged by elemental category (i.e., starting from the top, alkali, alkaline, 3d TM, 4d TM, metalloid, post-TM and C) followed by ascending atomic numbers in each elemental category.}
\label{fig:feature_imp}
\end{figure}

The normalized spectral regional feature importance of all elements in predicting certain CN is shown in Figure \ref{fig:feature_imp}. The x-axis denotes the CN grouped by the spectral region as shown in labels on the top of the graph and the y-axis shows the elements grouped by their corresponding elemental groups. For elements that do not have certain CNs, the feature importance is set to 0. Unsurprisingly, the ``Pre+Main'' region of the features plays a key role in all corresponding CNs and in general, two spectral regions have higher feature importance than single ones. The high feature importance for joint spectral regions implies that full spectral characteristics are necessary, consistent with previous studies\cite{carboneClassificationLocalChemical2019}. Even for CN4, the highest feature importance is achieved using ``Pre+Main" spectral regions followed by ``Pre+Post''. In addition, ``Main+Post'' becomes more important with increasing CN, in good agreement with previous studies\cite{yamamotoAssignmentPreedgePeaks2008, fargesTiEdgeXANES1997, carboneClassificationLocalChemical2019}.

For the first-row ($3d$) transition metals, the pre-edge plays an important role. This is due to the well-known fact that $3d$ transition metals with tetrahedral geometries tend to have strong pre-edge intensity due to the hybridization of unoccupied $p$ and $d$ states\cite{cottonSoftRayAbsorption1956,cottonSoftRayAbsorption1956a}. In addition, the early $3d$ transition metals tend to have stronger pre-edge effects than late ones. Our data-driven approach is able to capture this relationship known from group theory analysis.

\section{Discussion}

In summary, we have demonstrated that random forest models trained on FEFF-computed K-edge XANES can be used to directly predict the coordination environment - CN \textit{and} CM - with high accuracy. In contrast to prior works, we eschew a rigid classification of coordination environments into mutually-exclusive labels, opting instead for a more rigorous, mathematical definition of coordination environment based on multiple labels with order parameters. 

Prior works to identify coordination environment from XANES have primarily focused on deep learning models, i.e., MLP and CNN.\cite{timoshenkoSupervisedMachineLearningBasedDetermination2017,carboneClassificationLocalChemical2019} While such deep learning models perform respectably, especially for transition metals, one major finding of our work is that the random forest models outperform them by significant margins. The likely reason is that deep learning models are notoriously data-hungry; for many elements, there is insufficient data to train such models properly. On the other hand, the accuracy of the optimized random forest models shows little/weak dependence on data size and much stronger dependence on label entropy, suggesting that the random forest models are limited by task difficulty as opposed to data limitations. 

We also strongly advocate the development of baseline models for the evaluation of ML models in materials science. The commonly-used ``accuracy'' metric means little without this context. As is evident from Figure \ref{fig:acc_improvement}, many elements, the $3d$ transition metals being a notable category, have very high probabilities of being in a particular coordination environment, i.e., information content/label entropy is low. For instance, any ML model that achieves anything less than 89\% accuracy in classifying the coordination environment of Ti is, in effect, under-performing relative to a trivial model that always identifies Ti as being in a six-coordinated octahedron environment. Indeed, the random forest models yield far more substantial accuracy improvements in coordination environment identification over the baseline model for elements that exist in many different environments with high probability, e.g., Cu, Zn, C and the alkali metals. For the $3d$ transition metals, the accuracy improvement is a relatively modest $\sim 20$\%, even though all instances achieved high absolute accuracies exceeding 90\%.

Finally and most importantly, we demonstrate a data-augmentation approach that enables the random forest models trained on computed data to be directly applied to experimental K-edge XANES with minimal loss in accuracy. The models achieved an outstanding accuracy of $\sim82.1\%$ in identifying the dominant coordination environment over a diverse experimental spectra test set comprising 28 experimental K-edge XANES spectra of 13 chemical species. This addresses a critical gap in ML-based K-edge XANES analysis. High-quality experimental XANES data is difficult and expensive to obtain, and high-throughput computations are currently the only approach to generate large and diverse XANES datasets. Being able to develop a coordination environment identification ML model using the latter that can be applied to the former is therefore of major value, and represents a transformative advance in the application of ML to coordination environment identification.

\section{Methods}

\subsection{Construction of coordination environment ranking labels}

Given a real-valued vector $\mathbf{\hat{OP}} \in \mathbb{R}^{L}$, the $i$-th OP represents how closely the site's local coordination environment resembles a CN condition or a specific CM. A threshold $t$ is applied to $\mathbf{\hat{OPs}}$ to create a bipartition of relevant and irrelevant CN and CM labels. The multi-label prediction $\mathbf{\hat{y}}$ can be obtained as:
\begin{equation}
    \hat{y}_{j} = 
    \begin{cases}
    1 & \text{if $\hat{OPs}_{j} \geq t$} \\
    0 & \text{if $\hat{OPs}_{j} < t$}.\\
    \end{cases}
\end{equation}
Instead of using an arbitrary threshold like 0.5, we adopted the concept of label cardinality (LCard) and calibrated the threshold $t$ to minimize the possibility of a spectrum being assigned to the no-label set. The LCard\cite{tsoumakasMultiLabelClassification2007} is a standard measure of ``multi-labeled-ness'', which is simply the average number of labels associated with each example. For $N$ examples and $L$ labels, the LCard measure can be calculated as:
\begin{equation}
\mathrm{LCard} = \frac{1}{N}\sum_{i=1}^{N}\sum_{j=1}^{L}y_{j}^{i}. 
\end{equation}

The threshold $t_{1}$ for CN and threshold $t_{2}$ for CM were calibrated using the same procedure as follows:
\begin{equation}
t = \underset{t}{\mathrm{argmin}}||\mathrm{LCard}(D_{site-specific})-\mathrm{LCard}(D_{site-averaged})||,
\end{equation}
where $D_{site-specific}$ and $D_{site-averaged}$ are the dataset of $\sim110,000$ site-specific and $\sim36,000$ site-averaged computed K-edge XANES spectra, respectively. The site-averaged spectral dataset was also considered here as experimentally measured XANES spectra are the averaged absorption coefficients. The OPs of site-averaged spectra were obtained by averaging site-specific OPs of the same element. The calibration procedure aims at minimizing the difference between label cardinality of site-specific spectra and that of site-averaged spectra. This calibration approach has been found to be more effective and efficient in reducing the probability of empty-set prediction issues.\cite{readClassifierChainsMultilabel2011} 

We evaluated the threshold value $t_{1}$ and $t_{2}$ from 0 to 0.4 at 0.01 intervals. The average number of CN labels associated with each spectrum dropped below 1 when $t_{1}$ exceeds 0.4, and this was set at the upper limit. For the CN label set, we found that the LCard difference between the site-specific dataset and site-averaged dataset is minimized at $t_{1} = 0.2$. The average number of CN labels associated with each spectral example was $\sim 1.2$. For the CM label set, the difference in LCard between the two datasets reaches a minimum at $t_{2} = 0.05$. The average number of coordination environment labels associated with each spectrum was $\sim 3.2$. 

After applying the calibrated thresholds, we then encoded the CN and CM label sets into the form of ranking labels in terms of descending OPs. Using 0.2 as cutoff for CN OPs, the average number of CN ranking labels per element was 10. Note that the labels contain joint labels such as CN4-CN6. In the CM classification task, the average number of CM ranking labels is 5 per element per CN. As expected, the distribution of relevant CN labels, i.e., CN with $p_{CN} \geq 0.2$, was inhomogeneous (Figure S2). For each element, there are a few dominant CNs with an order of magnitude more data points than the other CNs. In the CM classification problem, we therefore restricted our consideration to those most abundant CN cases of each elemental group. Only CN $\leq 8$ were considered for the CM classification task as no target CM was provided for CN = 9-11 and only one CM was provided for CN = 12.

For each absorbing specie, we excluded CN and CM ranking labels with less than 30 samples. After applying this rule, all Tc, Ru and Rh ions are six coordinated. Therefore, we removed the their K-edge XANES from the first step CN classification task's training dataset. For the CM classification task, we repeated this operation and excluded those sub-datasets (see Table S1) associated with only one CM label from the training dataset as well. The final CNs in each elemental group that were subject the coordination environment classification task are given in Table \ref{tab:elementcncmclassification}. 

\begin{table}[]
    \centering
    \begin{tabular}{c|c}
         Element Group & CN \\
         \hline
         Alkali & 3-8\\
         Alkaline & 4-8\\
         Metalloid & 3-4\\
         Carbon & 2-4\\
         Transition metal & 4-6\\
         Post-transition metal & 4-6\\
    \end{tabular}
    \caption{Coordination number (CN) for each elemental group that were subject to coordination environment classification task.}
    \label{tab:elementcncmclassification}
\end{table}

To validate the necessity of using ranking labels to represent the absorption elements' coordination environments, we visualized the joint distributions of the CN and CM OPs of the alkali and the transition metal elemental group (Figure S7). From Figure S7, we observe that there are correlations across different CN OPs or CM OPs and multiple coordination environments coexist. We also note that the correlation between CM OPs is quite substantial and most six-coordinated transition metal ions' coordination patterns resemble two or more CMs with OPs exceeding $\geq 0.4$. These findings emphasize that labeling the absorbing sites' coordination environments with one label cannot adequately represent the full coordination environment. 

\subsection{Hyper-parameter optimization of machine learning algorithms}

In this work, we use the top-1 accuracy and Jaccard index as metrics to evaluate the performance of classifiers. The top-1 accuracy of a classifier is evaluated by its ability to yield the top-ranked coordination environment. The Jaccard index measures the overlaps between the true CN-CM labels and the predicted CN-CM labels. Let $y_{n}$ be the ground true CN-CM label set and $\hat y_{n}$ be the predicted CN-CM label made by a classifier. The Jaccard index can be computed based on the number of labels in the intersection set divided by the number of labels in the union set. 
\begin{equation}
J(y_{n}, \hat{y}_{n}) = \frac{|y_{n}\cap \hat{y}_{n}|}{|y_{n}\cup\hat{y}_{n}|}. 
\end{equation}

The Jaccard index yields a number (0-100\%) indicating how well a given classifier identifies all relevant coordination environments compared to the correct coordination environments. 

The hyper-parameter space investigated for each ML model are as follows: 
\begin{enumerate}
    \item $k$NN: The $k$-nearest neighbors classifier was optimized with respect to the number of neighbors ($N$) and the distance metric ($p$). The values of $N$ examined were 10, 20, 30, and 50. The minimum value of $N = 10$ was set to avoid overfitting and increase the generalizability of models. The Manhattan distance and Euclidean distance were used to assess the distances metric effects.
    \item Random forest classifier: The number of trees in the forest was tested at values 10, 20, 30, 50, 100, 200. The rest of the parameters were kept at the default settings.
    \item Multi-layer perceptron (MLP): For the MLP classifier, the number of hidden layers ($L$) was varied from 1 to 3, and the number of neurons in each hidden layer was varied from 10 to 100. The activation functions tested were the logistic, tanh and ReLU functions. 
    \item Support vector classifier (SVC): The penalty parameter $C$ was drawn exponentially from 0.001 to 100.0. The maximum value of $C$ was set at 100.0, as high $C$ is prone to overfitting. Two kernel coefficient ($\gamma$) values were tested: (a) 1 divided by the number of features ($\gamma = 0.005$), and (b) 1 divided by the number of features multiplied by the variance of the spectral absorption coefficients ($\gamma \simeq 0.013$). The radial basis function (RBF) kernel was set as the number of observations is one to two orders of magnitude higher than the number of features in the training data. In addition, a previous study\cite{keerthiAsymptoticBehaviorsSupport2003} has shown that it is unnecessary to consider the linear kernel if the model selection is conducted using the RBF kernel. 
    \item Convolution neural network (CNN): The two-layer CNN classifier was used. The two layers were fully connected, with feed-forward hidden layers with 50 and 100 neurons, ending with a softmax output layer. The number of neurons in the output layer equals to the number of target ranking labels. 
\end{enumerate}

For CN ranking labels classification, we found that the model using 10 nearest neighbors and Manhattan distance performs the best for $k$NN models.  The random forest classifier's performance converged at 30 trees for all elemental groups. For the MLP classifier, the two-layer neural network architecture with ReLU activation function outperformed the rest of the models with tanh or logistic sigmoid neurons. The best MLP model had 50 neurons in the first hidden layer and 100 in the second hidden layer. We found that further increasing number of hidden layers has a detrimental effect on classification performance. For the RBF SVC classifier, the model with $C=100$ and $\gamma \simeq 0.013$ performed the best.

For the CM ranking labels classification task. The optimum CN classifiers' parameter configurations were the best sets for $k$NN classifier, MLP classifier, and RBF SVC classifier as well. We found that the random forest classifier performed the best when the number of trees in the forest equals 50.

\begin{acknowledgement}

This work was intellectually led by the Data Infrastructure Building Blocks (DIBBS) Local Spectroscopy Data Infrastructure (LSDI) project funded by National Science Foundation (NSF), under Award Number 1640899. The authors thank Kristin A. Persson, Shyam Dwaraknath, J. Rehr and A. Dozier for helpful discussions. Computational resources were provided by the NSF DIBBs funding as well as the Triton Shared Computing Cluster (TSCC) at the University of California, San Diego. 

\end{acknowledgement}

\section{Author contributions}
S.P.O, C.Z. and C.C. proposed the concept. C.Z. and C.C. carried out the calculations and analysis with the help of Y.M.C. and S.P.O. C.Z. and C.C. prepared the initial draft of the manuscript. All authors contributed to the discussions and revisions of the manuscript.

\section{Declaration of interests}

The authors declare no competing interests.

\section{Data and code availability}

The K-edge XANES data is available from Materials Project website under the XAS app (\url{https://materialsproject.org/#apps/xas/}). The models presented in this work are available in the \textit{veidt} python package (\url{https://github.com/materialsvirtuallab/veidt}), developed by the Materials Virtual Lab.

\section{Additional information}

\textbf{Supplementary Information} accompanies this paper at 

\bibliography{mainbib}

\providecommand{\latin}[1]{#1}
\makeatletter
\providecommand{\doi}
  {\begingroup\let\do\@makeother\dospecials
  \catcode`\{=1 \catcode`\}=2 \doi@aux}
\providecommand{\doi@aux}[1]{\endgroup\texttt{#1}}
\makeatother
\providecommand*\mcitethebibliography{\thebibliography}
\csname @ifundefined\endcsname{endmcitethebibliography}
  {\let\endmcitethebibliography\endthebibliography}{}
\begin{mcitethebibliography}{54}
\providecommand*\natexlab[1]{#1}
\providecommand*\mciteSetBstSublistMode[1]{}
\providecommand*\mciteSetBstMaxWidthForm[2]{}
\providecommand*\mciteBstWouldAddEndPuncttrue
  {\def\EndOfBibitem{\unskip.}}
\providecommand*\mciteBstWouldAddEndPunctfalse
  {\let\EndOfBibitem\relax}
\providecommand*\mciteSetBstMidEndSepPunct[3]{}
\providecommand*\mciteSetBstSublistLabelBeginEnd[3]{}
\providecommand*\EndOfBibitem{}
\mciteSetBstSublistMode{f}
\mciteSetBstMaxWidthForm{subitem}{(\alph{mcitesubitemcount})}
\mciteSetBstSublistLabelBeginEnd
  {\mcitemaxwidthsubitemform\space}
  {\relax}
  {\relax}

\bibitem[O'Day \latin{et~al.}(2000)O'Day, Newville, Neuhoff, Sahai, and
  Carroll]{odayXRayAbsorptionSpectroscopy2000}
O'Day,~P.~A.; Newville,~M.; Neuhoff,~P.~S.; Sahai,~N.; Carroll,~S.~A.
  \emph{Journal of Colloid and Interface Science} \textbf{2000}, \emph{222},
  184--197\relax
\mciteBstWouldAddEndPuncttrue
\mciteSetBstMidEndSepPunct{\mcitedefaultmidpunct}
{\mcitedefaultendpunct}{\mcitedefaultseppunct}\relax
\EndOfBibitem
\bibitem[Chaurand \latin{et~al.}(2007)Chaurand, Rose, Briois, Salome, Proux,
  Nassif, Olivi, Susini, Hazemann, and
  Bottero]{chaurandNewMethodologicalApproach2007}
Chaurand,~P.; Rose,~J.; Briois,~V.; Salome,~M.; Proux,~O.; Nassif,~V.;
  Olivi,~L.; Susini,~J.; Hazemann,~J.-l.; Bottero,~J.-y. \emph{The Journal of
  Physical Chemistry B} \textbf{2007}, \emph{111}, 5101--5110\relax
\mciteBstWouldAddEndPuncttrue
\mciteSetBstMidEndSepPunct{\mcitedefaultmidpunct}
{\mcitedefaultendpunct}{\mcitedefaultseppunct}\relax
\EndOfBibitem
\bibitem[Silversmit \latin{et~al.}(2005)Silversmit, Bokhoven, Poelman, Eerden,
  and Marin]{silversmitStructureSupportedUnsupported2005}
Silversmit,~G.; Bokhoven,~J. A.~V.; Poelman,~H.; Eerden,~A. M. J. V.~D.;
  Marin,~G.~B. \textbf{2005}, \emph{285}, 151--162\relax
\mciteBstWouldAddEndPuncttrue
\mciteSetBstMidEndSepPunct{\mcitedefaultmidpunct}
{\mcitedefaultendpunct}{\mcitedefaultseppunct}\relax
\EndOfBibitem
\bibitem[Farges \latin{et~al.}(1997)Farges, Brown, and
  Rehr]{fargesTiEdgeXANES1997}
Farges,~F.; Brown,~G.~E.; Rehr,~J.~J. \emph{Physical Review B} \textbf{1997},
  \emph{56}, 1809--1819\relax
\mciteBstWouldAddEndPuncttrue
\mciteSetBstMidEndSepPunct{\mcitedefaultmidpunct}
{\mcitedefaultendpunct}{\mcitedefaultseppunct}\relax
\EndOfBibitem
\bibitem[Farges \latin{et~al.}(2001)Farges, Brown, Petit, and
  Munoz]{fargesTransitionElementsWaterbearing2001}
Farges,~F.; Brown,~G.~E.; Petit,~P.-E.; Munoz,~M. \emph{Geochimica et
  Cosmochimica Acta} \textbf{2001}, \emph{65}, 1665--1678\relax
\mciteBstWouldAddEndPuncttrue
\mciteSetBstMidEndSepPunct{\mcitedefaultmidpunct}
{\mcitedefaultendpunct}{\mcitedefaultseppunct}\relax
\EndOfBibitem
\bibitem[DeBeer~George \latin{et~al.}(2005)DeBeer~George, Brant, and
  Solomon]{debeergeorgeMetalLigandKEdge2005}
DeBeer~George,~S.; Brant,~P.; Solomon,~E.~I. \emph{Journal of the American
  Chemical Society} \textbf{2005}, \emph{127}, 667--674\relax
\mciteBstWouldAddEndPuncttrue
\mciteSetBstMidEndSepPunct{\mcitedefaultmidpunct}
{\mcitedefaultendpunct}{\mcitedefaultseppunct}\relax
\EndOfBibitem
\bibitem[Westre \latin{et~al.}(1997)Westre, Kennepohl, DeWitt, Hedman, Hodgson,
  and Solomon]{westreMultipletAnalysisFe1997}
Westre,~T.~E.; Kennepohl,~P.; DeWitt,~J.~G.; Hedman,~B.; Hodgson,~K.~O.;
  Solomon,~E.~I. \emph{Journal of the American Chemical Society} \textbf{1997},
  \emph{119}, 6297--6314\relax
\mciteBstWouldAddEndPuncttrue
\mciteSetBstMidEndSepPunct{\mcitedefaultmidpunct}
{\mcitedefaultendpunct}{\mcitedefaultseppunct}\relax
\EndOfBibitem
\bibitem[Yamamoto(2008)]{yamamotoAssignmentPreedgePeaks2008}
Yamamoto,~T. \emph{X-Ray Spectrometry} \textbf{2008}, \emph{37}, 572--584\relax
\mciteBstWouldAddEndPuncttrue
\mciteSetBstMidEndSepPunct{\mcitedefaultmidpunct}
{\mcitedefaultendpunct}{\mcitedefaultseppunct}\relax
\EndOfBibitem
\bibitem[Sano \latin{et~al.}(1992)Sano, Komorita, and
  Yamatera]{sanoXANESSpectraCopper1992}
Sano,~M.; Komorita,~S.; Yamatera,~H. \emph{Inorganic Chemistry} \textbf{1992},
  \emph{31}, 459--463\relax
\mciteBstWouldAddEndPuncttrue
\mciteSetBstMidEndSepPunct{\mcitedefaultmidpunct}
{\mcitedefaultendpunct}{\mcitedefaultseppunct}\relax
\EndOfBibitem
\bibitem[Farges \latin{et~al.}(2009)Farges, Jr, and
  Xanes]{fargesPreedgeAnalysisMn2009}
Farges,~E. C. {\AE}.~F.; Jr,~{\AE}. G. E.~B.; Xanes,~M. {\'A}. G.~{\'A}.
  \textbf{2009}, 111--126\relax
\mciteBstWouldAddEndPuncttrue
\mciteSetBstMidEndSepPunct{\mcitedefaultmidpunct}
{\mcitedefaultendpunct}{\mcitedefaultseppunct}\relax
\EndOfBibitem
\bibitem[{Fern{\'a}ndez-Garc{\'i}a}(2002)]{fernandez-garciaXANESAnalysisCatalytic2002}
{Fern{\'a}ndez-Garc{\'i}a},~M. \emph{Catalysis Reviews: Science and
  Engineering} \textbf{2002}, \emph{44}, 59--121\relax
\mciteBstWouldAddEndPuncttrue
\mciteSetBstMidEndSepPunct{\mcitedefaultmidpunct}
{\mcitedefaultendpunct}{\mcitedefaultseppunct}\relax
\EndOfBibitem
\bibitem[Manceau \latin{et~al.}(2014)Manceau, Marcus, and
  Lenoir]{manceauEstimatingNumberPure2014}
Manceau,~A.; Marcus,~M.; Lenoir,~T. \emph{Journal of Synchrotron Radiation}
  \textbf{2014}, \emph{21}, 1140--1147\relax
\mciteBstWouldAddEndPuncttrue
\mciteSetBstMidEndSepPunct{\mcitedefaultmidpunct}
{\mcitedefaultendpunct}{\mcitedefaultseppunct}\relax
\EndOfBibitem
\bibitem[Fay \latin{et~al.}(1992)Fay, Proctor, Hoffmann, Houalla, and
  Hercules]{fayDeterminationMoSurface1992}
Fay,~M.~J.; Proctor,~A.; Hoffmann,~D.~P.; Houalla,~M.; Hercules,~D.~M.
  \emph{Mikrochimica Acta} \textbf{1992}, \emph{109}, 281--293\relax
\mciteBstWouldAddEndPuncttrue
\mciteSetBstMidEndSepPunct{\mcitedefaultmidpunct}
{\mcitedefaultendpunct}{\mcitedefaultseppunct}\relax
\EndOfBibitem
\bibitem[Beauchemin \latin{et~al.}(2002)Beauchemin, Hesterberg, and
  Beauchemin]{beaucheminPrincipalComponentAnalysis2002}
Beauchemin,~S.; Hesterberg,~D.; Beauchemin,~M. \emph{Soil Science Society of
  America Journal} \textbf{2002}, \emph{66}, 83\relax
\mciteBstWouldAddEndPuncttrue
\mciteSetBstMidEndSepPunct{\mcitedefaultmidpunct}
{\mcitedefaultendpunct}{\mcitedefaultseppunct}\relax
\EndOfBibitem
\bibitem[Bajt \latin{et~al.}(1994)Bajt, Sutton, and
  Delaney]{bajtXrayMicroprobeAnalysis1994}
Bajt,~S.; Sutton,~S.; Delaney,~J. \emph{Geochimica et Cosmochimica Acta}
  \textbf{1994}, \emph{58}, 5209--5214\relax
\mciteBstWouldAddEndPuncttrue
\mciteSetBstMidEndSepPunct{\mcitedefaultmidpunct}
{\mcitedefaultendpunct}{\mcitedefaultseppunct}\relax
\EndOfBibitem
\bibitem[Tanaka and Mizoguchi(2009)Tanaka, and
  Mizoguchi]{tanakaFirstprinciplesCalculationsXray2009}
Tanaka,~I.; Mizoguchi,~T. \emph{Journal of Physics: Condensed Matter}
  \textbf{2009}, \emph{21}, 104201\relax
\mciteBstWouldAddEndPuncttrue
\mciteSetBstMidEndSepPunct{\mcitedefaultmidpunct}
{\mcitedefaultendpunct}{\mcitedefaultseppunct}\relax
\EndOfBibitem
\bibitem[Rehr and Albers(2000)Rehr, and
  Albers]{rehrTheoreticalApproachesXray2000}
Rehr,~J.~J.; Albers,~R.~C. \emph{Reviews of Modern Physics} \textbf{2000},
  \emph{72}, 621--654\relax
\mciteBstWouldAddEndPuncttrue
\mciteSetBstMidEndSepPunct{\mcitedefaultmidpunct}
{\mcitedefaultendpunct}{\mcitedefaultseppunct}\relax
\EndOfBibitem
\bibitem[Rehr \latin{et~al.}(2010)Rehr, Kas, Vila, Prange, and
  Jorissen]{rehrParameterfreeCalculationsXray2010}
Rehr,~J.~J.; Kas,~J.~J.; Vila,~F.~D.; Prange,~M.~P.; Jorissen,~K.
  \emph{Physical Chemistry Chemical Physics} \textbf{2010}, \emph{12},
  5503\relax
\mciteBstWouldAddEndPuncttrue
\mciteSetBstMidEndSepPunct{\mcitedefaultmidpunct}
{\mcitedefaultendpunct}{\mcitedefaultseppunct}\relax
\EndOfBibitem
\bibitem[Laskowski and Blaha(2010)Laskowski, and
  Blaha]{laskowskiUnderstandingXrayAbsorption2010}
Laskowski,~R.; Blaha,~P. \emph{Physical Review B} \textbf{2010}, \emph{82},
  205104\relax
\mciteBstWouldAddEndPuncttrue
\mciteSetBstMidEndSepPunct{\mcitedefaultmidpunct}
{\mcitedefaultendpunct}{\mcitedefaultseppunct}\relax
\EndOfBibitem
\bibitem[Zheng \latin{et~al.}(2018)Zheng, Mathew, Chen, Chen, Tang, Dozier,
  Kas, Vila, Rehr, Piper, Persson, and
  Ong]{zhengAutomatedGenerationEnsemblelearned2018}
Zheng,~C.; Mathew,~K.; Chen,~C.; Chen,~Y.; Tang,~H.; Dozier,~A.; Kas,~J.~J.;
  Vila,~F.~D.; Rehr,~J.~J.; Piper,~L. F.~J.; Persson,~K.~A.; Ong,~S.~P.
  \emph{npj Computational Materials} \textbf{2018}, \emph{4}, 12\relax
\mciteBstWouldAddEndPuncttrue
\mciteSetBstMidEndSepPunct{\mcitedefaultmidpunct}
{\mcitedefaultendpunct}{\mcitedefaultseppunct}\relax
\EndOfBibitem
\bibitem[Mathew \latin{et~al.}(2018)Mathew, Zheng, Winston, Chen, Dozier, Rehr,
  Ong, and Persson]{mathewHighthroughputComputationalXray2018}
Mathew,~K.; Zheng,~C.; Winston,~D.; Chen,~C.; Dozier,~A.; Rehr,~J.~J.;
  Ong,~S.~P.; Persson,~K.~A. \emph{Scientific Data} \textbf{2018}, \emph{5},
  180151\relax
\mciteBstWouldAddEndPuncttrue
\mciteSetBstMidEndSepPunct{\mcitedefaultmidpunct}
{\mcitedefaultendpunct}{\mcitedefaultseppunct}\relax
\EndOfBibitem
\bibitem[Jain \latin{et~al.}(2013)Jain, Ong, Hautier, Chen, Richards, Dacek,
  Cholia, Gunter, Skinner, Ceder, and
  a.~Persson]{jainCommentaryMaterialsProject2013}
Jain,~A.; Ong,~S.~P.; Hautier,~G.; Chen,~W.; Richards,~W.~D.; Dacek,~S.;
  Cholia,~S.; Gunter,~D.; Skinner,~D.; Ceder,~G.; a.~Persson,~K. \emph{APL
  Materials} \textbf{2013}, \emph{1}, 011002\relax
\mciteBstWouldAddEndPuncttrue
\mciteSetBstMidEndSepPunct{\mcitedefaultmidpunct}
{\mcitedefaultendpunct}{\mcitedefaultseppunct}\relax
\EndOfBibitem
\bibitem[Timoshenko \latin{et~al.}(2017)Timoshenko, Lu, Lin, and
  Frenkel]{timoshenkoSupervisedMachineLearningBasedDetermination2017}
Timoshenko,~J.; Lu,~D.; Lin,~Y.; Frenkel,~A.~I. \emph{Journal of Physical
  Chemistry Letters} \textbf{2017}, \emph{8}, 5091--5098\relax
\mciteBstWouldAddEndPuncttrue
\mciteSetBstMidEndSepPunct{\mcitedefaultmidpunct}
{\mcitedefaultendpunct}{\mcitedefaultseppunct}\relax
\EndOfBibitem
\bibitem[Carbone \latin{et~al.}(2019)Carbone, Yoo, Topsakal, and
  Lu]{carboneClassificationLocalChemical2019}
Carbone,~M.~R.; Yoo,~S.; Topsakal,~M.; Lu,~D. \emph{Physical Review Materials}
  \textbf{2019}, \emph{3}, 033604\relax
\mciteBstWouldAddEndPuncttrue
\mciteSetBstMidEndSepPunct{\mcitedefaultmidpunct}
{\mcitedefaultendpunct}{\mcitedefaultseppunct}\relax
\EndOfBibitem
\bibitem[Kiyohara \latin{et~al.}(2018)Kiyohara, Miyata, Tsuda, and
  Mizoguchi]{kiyoharaDatadrivenApproachPrediction2018}
Kiyohara,~S.; Miyata,~T.; Tsuda,~K.; Mizoguchi,~T. \emph{Scientific Reports}
  \textbf{2018}, \emph{8}, 13548\relax
\mciteBstWouldAddEndPuncttrue
\mciteSetBstMidEndSepPunct{\mcitedefaultmidpunct}
{\mcitedefaultendpunct}{\mcitedefaultseppunct}\relax
\EndOfBibitem
\bibitem[Suzuki \latin{et~al.}(2019)Suzuki, Hino, Kotsugi, and
  Ono]{suzukiAutomatedEstimationMaterials2019}
Suzuki,~Y.; Hino,~H.; Kotsugi,~M.; Ono,~K. \emph{npj Computational Materials}
  \textbf{2019}, \emph{5}, 39\relax
\mciteBstWouldAddEndPuncttrue
\mciteSetBstMidEndSepPunct{\mcitedefaultmidpunct}
{\mcitedefaultendpunct}{\mcitedefaultseppunct}\relax
\EndOfBibitem
\bibitem[Ankudinov \latin{et~al.}(1998)Ankudinov, Ravel, Rehr, and
  Conradson]{ankudinovRealspaceMultiplescatteringCalculation1998}
Ankudinov,~A.~L.; Ravel,~B.; Rehr,~J.~J.; Conradson,~S.~D. \emph{Physical
  Review B} \textbf{1998}, \emph{58}, 7565--7576\relax
\mciteBstWouldAddEndPuncttrue
\mciteSetBstMidEndSepPunct{\mcitedefaultmidpunct}
{\mcitedefaultendpunct}{\mcitedefaultseppunct}\relax
\EndOfBibitem
\bibitem[Perdew \latin{et~al.}(1996)Perdew, Burke, and
  Ernzerhof]{perdewGeneralizedGradientApproximation1996}
Perdew,~J.~P.; Burke,~K.; Ernzerhof,~M. \emph{Physical Review Letters}
  \textbf{1996}, \emph{77}, 3865--3868\relax
\mciteBstWouldAddEndPuncttrue
\mciteSetBstMidEndSepPunct{\mcitedefaultmidpunct}
{\mcitedefaultendpunct}{\mcitedefaultseppunct}\relax
\EndOfBibitem
\bibitem[Zimmermann \latin{et~al.}(2017)Zimmermann, Horton, Jain, and
  Haranczyk]{zimmermannAssessingLocalStructure2017}
Zimmermann,~N. E.~R.; Horton,~M.~K.; Jain,~A.; Haranczyk,~M. \emph{Frontiers in
  Materials} \textbf{2017}, \emph{4}, 1--13\relax
\mciteBstWouldAddEndPuncttrue
\mciteSetBstMidEndSepPunct{\mcitedefaultmidpunct}
{\mcitedefaultendpunct}{\mcitedefaultseppunct}\relax
\EndOfBibitem
\bibitem[Ong \latin{et~al.}(2013)Ong, Richards, Jain, Hautier, Kocher, Cholia,
  Gunter, Chevrier, a.~Persson, and Ceder]{ongPythonMaterialsGenomics2013}
Ong,~S.~P.; Richards,~W.~D.; Jain,~A.; Hautier,~G.; Kocher,~M.; Cholia,~S.;
  Gunter,~D.; Chevrier,~V.~L.; a.~Persson,~K.; Ceder,~G. \emph{Computational
  Materials Science} \textbf{2013}, \emph{68}, 314--319\relax
\mciteBstWouldAddEndPuncttrue
\mciteSetBstMidEndSepPunct{\mcitedefaultmidpunct}
{\mcitedefaultendpunct}{\mcitedefaultseppunct}\relax
\EndOfBibitem
\bibitem[Ward \latin{et~al.}(2018)Ward, Dunn, Faghaninia, Zimmermann, Bajaj,
  Wang, Montoya, Chen, Bystrom, Dylla, Chard, Asta, Persson, Snyder, Foster,
  and Jain]{wardMatminerOpenSource2018}
Ward,~L. \latin{et~al.}  \emph{Computational Materials Science} \textbf{2018},
  \emph{152}, 60--69\relax
\mciteBstWouldAddEndPuncttrue
\mciteSetBstMidEndSepPunct{\mcitedefaultmidpunct}
{\mcitedefaultendpunct}{\mcitedefaultseppunct}\relax
\EndOfBibitem
\bibitem[Newville(2014)]{newvilleFundamentalsXAFS2014}
Newville,~M. \emph{Reviews in Mineralogy and Geochemistry} \textbf{2014},
  \emph{78}, 33--74\relax
\mciteBstWouldAddEndPuncttrue
\mciteSetBstMidEndSepPunct{\mcitedefaultmidpunct}
{\mcitedefaultendpunct}{\mcitedefaultseppunct}\relax
\EndOfBibitem
\bibitem[Read \latin{et~al.}(2011)Read, Pfahringer, Holmes, and
  Frank]{readClassifierChainsMultilabel2011}
Read,~J.; Pfahringer,~B.; Holmes,~G.; Frank,~E. \emph{Machine Learning}
  \textbf{2011}, \emph{85}, 333--359\relax
\mciteBstWouldAddEndPuncttrue
\mciteSetBstMidEndSepPunct{\mcitedefaultmidpunct}
{\mcitedefaultendpunct}{\mcitedefaultseppunct}\relax
\EndOfBibitem
\bibitem[Chang and Lin(2011)Chang, and Lin]{changLIBSVMLibrarySupport2011}
Chang,~C.-C.; Lin,~C.-J. \emph{ACM Trans. Intell. Syst. Technol.}
  \textbf{2011}, \emph{2}, 27:1--27:27\relax
\mciteBstWouldAddEndPuncttrue
\mciteSetBstMidEndSepPunct{\mcitedefaultmidpunct}
{\mcitedefaultendpunct}{\mcitedefaultseppunct}\relax
\EndOfBibitem
\bibitem[Breiman(2001)]{breimanRandomForests2001}
Breiman,~L. \emph{Machine Learning} \textbf{2001}, \emph{45}, 5--32\relax
\mciteBstWouldAddEndPuncttrue
\mciteSetBstMidEndSepPunct{\mcitedefaultmidpunct}
{\mcitedefaultendpunct}{\mcitedefaultseppunct}\relax
\EndOfBibitem
\bibitem[Pedregosa \latin{et~al.}(2011)Pedregosa, Varoquaux, Gramfort, Michel,
  Thirion, Grisel, Blondel, Prettenhofer, Weiss, and
  Dubourg]{pedregosaScikitlearnMachineLearning2011}
Pedregosa,~F.; Varoquaux,~G.; Gramfort,~A.; Michel,~V.; Thirion,~B.;
  Grisel,~O.; Blondel,~M.; Prettenhofer,~P.; Weiss,~R.; Dubourg,~V.
  \emph{Journal of machine learning research} \textbf{2011}, \emph{12},
  2825--2830\relax
\mciteBstWouldAddEndPuncttrue
\mciteSetBstMidEndSepPunct{\mcitedefaultmidpunct}
{\mcitedefaultendpunct}{\mcitedefaultseppunct}\relax
\EndOfBibitem
\bibitem[Waroquiers \latin{et~al.}(2017)Waroquiers, Gonze, Rignanese,
  {Welker-Nieuwoudt}, Rosowski, G{\"o}bel, Schenk, Degelmann, Andr{\'e}, Glaum,
  and Hautier]{waroquiersStatisticalAnalysisCoordination2017}
Waroquiers,~D.; Gonze,~X.; Rignanese,~G.-M.; {Welker-Nieuwoudt},~C.;
  Rosowski,~F.; G{\"o}bel,~M.; Schenk,~S.; Degelmann,~P.; Andr{\'e},~R.;
  Glaum,~R.; Hautier,~G. \emph{Chemistry of Materials} \textbf{2017},
  \emph{29}, 8346--8360\relax
\mciteBstWouldAddEndPuncttrue
\mciteSetBstMidEndSepPunct{\mcitedefaultmidpunct}
{\mcitedefaultendpunct}{\mcitedefaultseppunct}\relax
\EndOfBibitem
\bibitem[LeCun \latin{et~al.}(2015)LeCun, Bengio, and
  Hinton]{lecunDeepLearning2015}
LeCun,~Y.; Bengio,~Y.; Hinton,~G. \emph{Nature} \textbf{2015}, \emph{521},
  436--444\relax
\mciteBstWouldAddEndPuncttrue
\mciteSetBstMidEndSepPunct{\mcitedefaultmidpunct}
{\mcitedefaultendpunct}{\mcitedefaultseppunct}\relax
\EndOfBibitem
\bibitem[LeCun \latin{et~al.}(2010)LeCun, Kavukcuoglu, and
  Farabet]{lecunConvolutionalNetworksApplications2010}
LeCun,~Y.; Kavukcuoglu,~K.; Farabet,~C. Convolutional Networks and Applications
  in Vision. Proceedings of 2010 {{IEEE International Symposium}} on
  {{Circuits}} and {{Systems}}. 2010; pp 253--256\relax
\mciteBstWouldAddEndPuncttrue
\mciteSetBstMidEndSepPunct{\mcitedefaultmidpunct}
{\mcitedefaultendpunct}{\mcitedefaultseppunct}\relax
\EndOfBibitem
\bibitem[Shannon(1948)]{shannonMathematicalTheoryCommunication1948}
Shannon,~C.~E. \emph{The Bell System Technical Journal} \textbf{1948},
  \emph{27}, 379--423\relax
\mciteBstWouldAddEndPuncttrue
\mciteSetBstMidEndSepPunct{\mcitedefaultmidpunct}
{\mcitedefaultendpunct}{\mcitedefaultseppunct}\relax
\EndOfBibitem
\bibitem[Prado and M.Flank(2005)Prado, and M.Flank]{pradoSodiumEdgeXANES2005}
Prado,~R.~J.; M.Flank,~A. \emph{Physica Scripta} \textbf{2005}, 165\relax
\mciteBstWouldAddEndPuncttrue
\mciteSetBstMidEndSepPunct{\mcitedefaultmidpunct}
{\mcitedefaultendpunct}{\mcitedefaultseppunct}\relax
\EndOfBibitem
\bibitem[XAS()]{XASSpectraLibrary}
{{XAS Spectra Library}}. https://cars.uchicago.edu/xaslib\relax
\mciteBstWouldAddEndPuncttrue
\mciteSetBstMidEndSepPunct{\mcitedefaultmidpunct}
{\mcitedefaultendpunct}{\mcitedefaultseppunct}\relax
\EndOfBibitem
\bibitem[Ewels \latin{et~al.}(2016)Ewels, Sikora, Serin, Ewels, and
  Lajaunie]{ewelsCompleteOverhaulElectron2016}
Ewels,~P.; Sikora,~T.; Serin,~V.; Ewels,~C.~P.; Lajaunie,~L. \emph{Microscopy
  and Microanalysis} \textbf{2016}, \emph{22}, 717--724\relax
\mciteBstWouldAddEndPuncttrue
\mciteSetBstMidEndSepPunct{\mcitedefaultmidpunct}
{\mcitedefaultendpunct}{\mcitedefaultseppunct}\relax
\EndOfBibitem
\bibitem[Rana \latin{et~al.}(2014)Rana, Glatthaar, Gesswein, Sharma, Binder,
  Chernikov, Schumacher, and Banhart]{ranaLocalStructuralChanges2014}
Rana,~J.; Glatthaar,~S.; Gesswein,~H.; Sharma,~N.; Binder,~J.~R.;
  Chernikov,~R.; Schumacher,~G.; Banhart,~J. \emph{Journal of Power Sources}
  \textbf{2014}, \emph{255}, 439--449\relax
\mciteBstWouldAddEndPuncttrue
\mciteSetBstMidEndSepPunct{\mcitedefaultmidpunct}
{\mcitedefaultendpunct}{\mcitedefaultseppunct}\relax
\EndOfBibitem
\bibitem[Rana \latin{et~al.}(2014)Rana, Kloepsch, Li, Scherb, Schumacher,
  Winter, and Banhart]{ranaStructuralIntegrityElectrochemical2014}
Rana,~J.; Kloepsch,~R.; Li,~J.; Scherb,~T.; Schumacher,~G.; Winter,~M.;
  Banhart,~J. \emph{Journal of Materials Chemistry A} \textbf{2014}, \emph{2},
  9099\relax
\mciteBstWouldAddEndPuncttrue
\mciteSetBstMidEndSepPunct{\mcitedefaultmidpunct}
{\mcitedefaultendpunct}{\mcitedefaultseppunct}\relax
\EndOfBibitem
\bibitem[Weng \latin{et~al.}(2005)Weng, Waldo, and
  {Penner-Hahn}]{wengMethodNormalizationXray2005}
Weng,~T.~C.; Waldo,~G.~S.; {Penner-Hahn},~J.~E. \emph{Journal of Synchrotron
  Radiation} \textbf{2005}, \emph{12}, 506--510\relax
\mciteBstWouldAddEndPuncttrue
\mciteSetBstMidEndSepPunct{\mcitedefaultmidpunct}
{\mcitedefaultendpunct}{\mcitedefaultseppunct}\relax
\EndOfBibitem
\bibitem[Wu and Cohen(2006)Wu, and Cohen]{wuMoreAccurateGeneralized2006}
Wu,~Z.; Cohen,~R.~E. \emph{Physical Review B - Condensed Matter and Materials
  Physics} \textbf{2006}, \emph{73}, 2--7\relax
\mciteBstWouldAddEndPuncttrue
\mciteSetBstMidEndSepPunct{\mcitedefaultmidpunct}
{\mcitedefaultendpunct}{\mcitedefaultseppunct}\relax
\EndOfBibitem
\bibitem[Haas \latin{et~al.}(2009)Haas, Tran, and
  Blaha]{haasCalculationLatticeConstant2009}
Haas,~P.; Tran,~F.; Blaha,~P. \emph{Physical Review B} \textbf{2009},
  \emph{79}, 085104\relax
\mciteBstWouldAddEndPuncttrue
\mciteSetBstMidEndSepPunct{\mcitedefaultmidpunct}
{\mcitedefaultendpunct}{\mcitedefaultseppunct}\relax
\EndOfBibitem
\bibitem[Strobl \latin{et~al.}(2007)Strobl, Boulesteix, Zeileis, and
  Hothorn]{stroblBiasRandomForest2007}
Strobl,~C.; Boulesteix,~A.~L.; Zeileis,~A.; Hothorn,~T. \emph{BMC
  Bioinformatics} \textbf{2007}, \emph{8}\relax
\mciteBstWouldAddEndPuncttrue
\mciteSetBstMidEndSepPunct{\mcitedefaultmidpunct}
{\mcitedefaultendpunct}{\mcitedefaultseppunct}\relax
\EndOfBibitem
\bibitem[Cotton and Ballhausen(1956)Cotton, and
  Ballhausen]{cottonSoftRayAbsorption1956}
Cotton,~F.~A.; Ballhausen,~C.~J. \emph{The Journal of Chemical Physics}
  \textbf{1956}, \emph{25}, 617--619\relax
\mciteBstWouldAddEndPuncttrue
\mciteSetBstMidEndSepPunct{\mcitedefaultmidpunct}
{\mcitedefaultendpunct}{\mcitedefaultseppunct}\relax
\EndOfBibitem
\bibitem[Cotton and Hanson(1956)Cotton, and
  Hanson]{cottonSoftRayAbsorption1956a}
Cotton,~F.~A.; Hanson,~H.~P. \emph{The Journal of Chemical Physics}
  \textbf{1956}, \emph{25}, 619--623\relax
\mciteBstWouldAddEndPuncttrue
\mciteSetBstMidEndSepPunct{\mcitedefaultmidpunct}
{\mcitedefaultendpunct}{\mcitedefaultseppunct}\relax
\EndOfBibitem
\bibitem[Tsoumakas and Katakis(2007)Tsoumakas, and
  Katakis]{tsoumakasMultiLabelClassification2007}
Tsoumakas,~G.; Katakis,~I. \emph{International Journal of Data Warehousing and
  Mining} \textbf{2007}, \emph{3}, 1--13\relax
\mciteBstWouldAddEndPuncttrue
\mciteSetBstMidEndSepPunct{\mcitedefaultmidpunct}
{\mcitedefaultendpunct}{\mcitedefaultseppunct}\relax
\EndOfBibitem
\bibitem[Keerthi and Lin(2003)Keerthi, and
  Lin]{keerthiAsymptoticBehaviorsSupport2003}
Keerthi,~S.~S.; Lin,~C.~J. \emph{Neural Computation} \textbf{2003}, \emph{15},
  1667--1689\relax
\mciteBstWouldAddEndPuncttrue
\mciteSetBstMidEndSepPunct{\mcitedefaultmidpunct}
{\mcitedefaultendpunct}{\mcitedefaultseppunct}\relax
\EndOfBibitem
\end{mcitethebibliography}

\end{document}